\documentclass[12pt,preprint,showpacs,preprintnumbers]{revtex4}
\usepackage{amsfonts}
\usepackage{amsmath}
\usepackage{amssymb}
\usepackage{hyperref}
\usepackage{color}
\usepackage{graphicx}
\usepackage{amssymb}
\usepackage{amsmath}
\usepackage{graphicx}
\usepackage{dcolumn}
\usepackage{bm}
\usepackage{epsfig}
\usepackage[T1]{fontenc}
\usepackage{ae,aecompl}

\begin{document}

\title{Polar coordinate lattice Boltzmann kinetic modeling of detonation phenomena}
\author{Chuandong Lin$^{1}$,
Aiguo Xu$^{2,3,4}$\footnote{Corresponding author. E-mail address: Xu\_Aiguo@iapcm.ac.cn},
Guangcai Zhang$^{2,4,5}$,
Yingjun Li$^{1}$\footnote{Corresponding author. E-mail address: lyj@aphy.iphy.ac.cn}}
\affiliation{$^1$ State Key Laboratory for GeoMechanics and Deep Underground Engineering,
China University of Mining and Technology, Beijing 100083, P.R.China \\
$^2$ National Key Laboratory of Computational Physics,
Institute of Applied Physics and Computational Mathematics, P. O. Box 8009-26, Beijing 100088, P.R.China \\
$^3$ Center for Applied Physics and Technology, MOE Key Center for High Energy Density Physics Simulations, College of Engineering, Peking University, Beijing 100871, P.R.China \\
$^4$ State Key Laboratory of Explosion Science and Technology, Beijing Institute of Technology, Beijing 100081, P.R.China \\
$^5$ State Key Laboratory of Theoretical Physics, Institute of Theoretical
Physics, Chinese Academy of Sciences,Beijing 100190, P.R.China
 }
\date{\today}

\begin{abstract}
A novel polar coordinate lattice Boltzmann kinetic model for detonation phenomena is presented and applied to investigate typical implosion and explosion processes. In this model, the change of discrete distribution function due to local chemical reaction is dynamically coupled into in the modified lattice Boltzmann equation, which could recovery the Navier-Stokes equations, including contribution of chemical reaction, via the Chapman-Enskog expansion. For the numerical investigations, the main focuses are the nonequilibrium behaviors in these processes. The system at the disc center is always in its thermodynamic equilibrium. The internal kinetic energies in different degrees freedoms around the detonation front do not coincide due to the fluid viscosity. They show the maximum difference at the inflexion point where the pressure has the largest spatial derivative. The dependence of the reaction rate on the pressure, influences of the shock strength and reaction rate on the departure amplitude of the system from its local thermodynamic equilibrium are probed.
\end{abstract}

\maketitle

\section{Introduction}

The rapid and violent form of combustion called detonation \cite{Fickett1979} propagates through detonation wave which is a shock wave with chemical reaction. Given a wide range of application in science and engineering, shock and detonation have always been of great concern in the field of science and technology, such as the research and analysis on coal and gas outburst mechanism \cite{Cheng2004,Li2010}. The detonation phenomena are widely used in the acceleration of various projectiles, mining technologies, depositing of coating to a surface or cleaning of equipment, etc. Early in 1899 and 1905, Chapmann \cite{Chapman1899} and Jouguet \cite{Jouguet1905} presented CJ theory. This theory assumes that detonation front is a strong discontinuous plane with chemical reaction which immediately completes as soon as the detonation wave passes. In 1940s, Zeldovich \cite{Zeldovich1940}, Neumann \cite{Neumann1942} and Doering \cite{Doering1943} presented the well-known ZND model. This model gives an important conclusion that there is von-Neumann-peak at detonation wave front. Reactant is firstly pre-compressed by shock wave, and there is a continuous reaction zone behind the shock wave. Physical quantities (density, temperature, pressure and velocity) reach maximum values within the reaction zone.

Although detonation has been studied for more than one century \cite{Bjerketvedt1997}, it remains an active area of research in both theoretical studies and numerical simulations \cite{XunKun2000} due to its practical importance \cite{Wang2012}. So far, all chemical reaction models are empirical or semi-empirical formulas \cite{Sun1995}, such as the Arrhenius kinetics, forest fire burn, two-step model, Cochran's rate function \cite{Cochran1979}, Lee-Tarver model \cite{LeeTarver1980}, etc. Selecting appropriate chemical reaction kinetics is very important for describing detonation phenomena under consideration. In this paper, we adopt Cochran's rate function for chemical reaction, which is one of the most physically justifiable models satisfying simulation and experimental results \cite{Cao1986,Zhao1989}.

In recent decades Lattice Boltzmann (LB) method has achieved great success in various fields of fluid dynamics\cite{Succi-Book,Succi_RMP,Succi_Sci,Succi_PRL2005,Succi_PRL2006A,Succi_PRL2006B,Succi_PRL2009%
,Yeomans_PRL1997,Yeomans_PRL2001,Yeomans_PRL2002,Yeomans_PRL2004A,Yeomans_PRL2004B,Yeomans_PRL2004C%
,Yeomans_PRL2007,Yeomans_PRL2013,ShanChen,SYChen,DXZhang,HPFang,Guozhaoli2013,ProgPhys2014,Dawson1993,Weimar1996,
ZhangRenliang2014,ChenShiyi1996}. LB modeling for combustion phenomena \cite{Succi1997,Filippova1998,Filippova2000JCP,Filippova2000CPC,Yu2002%
,Yamamoto2002,Yamamoto2003,Yamamoto2005,Lee2006,Chiavazzo2010%
,ChenSheng2007,ChenSheng2008,ChenSheng2009,ChenSheng2010I,ChenSheng2010II,ChenSheng2010III,ChenSheng2011,ChenSheng2012} has been an interesting topic from early days.
In 1997, Succi et al. \cite{Succi1997} proposed the pioneering LB model for combustion systems under the assumptions of fast chemistry and cold flames with weak heat release.
In 1998 and 2000, Filippova and H\"{a}nel \cite{Filippova1998,Filippova2000JCP,Filippova2000CPC} proposed a kind of hybrid scheme for low Mach number reactive flows. The flow field is solved by modified lattice-BGK model and the transport equations for energy and species are solved by a finite difference scheme. In 2002, Yu et al. \cite{Yu2002} simulated scalar mixing in a multi-component flow and a chemical reacting flow using the LB method. In the same year, Yamamoto et al \cite{Yamamoto2002} presented a LB model for simulation of combustion, which includes reaction, diffusion and convection. In 2006, Lee et al. presented a new two-distribution LB equation algorithm to solve the laminar diffusion flames within the context of Burke-Schumann flame sheet model. In 2007, Chen et al \cite{ChenSheng2007} developed a novel coupled lattice Boltzmann model for two- and three-dimensional low Mach number combustion simulations. In this model, the fluid density can bear sharp changes. In the following year, another LB model was proposed for simulating combustion in two-dimensional system by Chen and and his cooperators \cite{ChenSheng2008}. Within this model the time step and the fluid particle speed can be adjusted dynamically. Later, based on their improved models, they presented a number of works \cite{ChenSheng2009,ChenSheng2010I,ChenSheng2010II,ChenSheng2010III,ChenSheng2011,ChenSheng2012}.

However, previous studies on LB model were mainly focused on isothermal and incompressible fluid systems. Those models generally can not recover the correct energy equation or describe enough the compressibility in the hydrodynamic limit, which makes difficult the modeling of systems with shock and/or detonation. At the same time, most of those LB models assume that exothermic reaction has no significant effect on fluid field, which also constrains the practical application of the models to most cases of combustion. In recent years, the development of LB models for high speed compressible flows \cite{Alexander1992,Alexander1993,Chen1994,McNamara1997,XuPan2007,XuGan,XuChen,Review2012} makes it possible to simulate systems with shock and detonation. Very recently Yan, Xu, Zhang, et al. proposed a Lattice Boltzmann Kinetic Model (LBKM) for detonation phenomena in Cartesian coordinates \cite{XuYan2013}. For simulating the explosion and implosion behaviors, a polar coordinate LB model is obviously more convenient. And there are many nice papers about LB formulations for axisymmetric flows in polar coordinates \cite{Halliday2001,Niu2003,Premnath2005,Reis2007,Guo2009}. In 2011 Watari \cite{Watari2011} proposed a finite-difference LB methods in polar coordinate system. Recently we \cite{XuLin2014} improved the LBKM by using a hybrid scheme so that it works also for supersonic flows. Within the improved model, the temporal evolution is calculated analytically and the convection term is solved via a Modified Warming-Beam (MWB) scheme. In this work, a new polar coordinate LBKM which is similar to and simpler than the one in Ref.\cite{XuLin2014} is used to study detonation phenomena.

In contrast to traditional methods based on Navier-Stokes description, the LBKM has some intrinsic superiority in describing kinetic mechanisms in systems where equilibrium and non-equilibrium behaviors coexist \cite{ProgPhys2014,Review2012,XuYan2013,XuLin2014}. The mini-review \cite{Review2012} presented a methodology to investigate non-equilibrium behaviors of the system by using the LB method. The non-equilibrium behaviors in various complex systems attract great attention \cite{XuYan2013,XuLin2014,XuGan2013,XuChen2013}. In the work \cite{XuYan2013} by Yan, Xu, Zhang, et al., some non-equilibrium behaviors around the von Neumann peak are obtained. In a recent work \cite{XuLin2014} we studied the non-equilibrium characteristics of the system around three kinds of interfaces, the shock wave, the rarefaction wave and the material interface, for two specific cases. We draw qualitative information on the actual distribution function. In this work, we further develop the LBKM with chemical reaction in polar coordinates to model the implosion and explosion phenomena and investigate the macroscopic behaviors due to deviating from local thermodynamic equilibrium in the detonation procedure.

The rest of the paper is structured as follows. In section II the polar coordinate LBKM for compressible fluid with chemical reaction is proposed for the first time. This model can recovery the Navier-Stokes equations with chemical reaction. The treatment of inner boundary around disc center is presented, and the manifestations of non-equilibrium characteristics are introduced. In section III we give the chemical reaction model and numerical verification, simulate implosion and explosion phenomena, and study the non-equilibrium characteristics of each case. The actual distribution functions around detonation wave are qualitatively illustrated. Section IV gives the conclusion and discussions.

\section{LBKM in polar coordinates}

\subsection{Modified Boltzmann equation with chemical reaction}

Here, for the purpose of simulating detonation which includes chemical reaction, we propose a novel Botlzmann equation, to the right side of which an artificial term $M$ is added. This term is called chemical term. The modified Boltzmann equation with the Bhatanger-Gross-Krook approximation reads
\begin{equation}
\frac{\partial f}{\partial t}+\mathbf{v}\cdot \nabla f = -\frac{1} {\tau }( f-f^{eq})+M.
\label{Boltzmann_chemical}
\end{equation}
where $f$ ($f^{eq}$) is the (equilibrium) distribution function; $\tau$ the relaxation time; $\mathbf{v}$ the discrete velocity.
To give the special form of $M$, the following assumptions are made:

1. The flow is describe by single function $f$. The the relaxation time $\tau$ is constant (not a function of density or temperature) and independent of time $t$ or space $\mathbf{r}$.

2. The flow is symmetric, and there are no external forces. The radiative heat loss is neglected.

3. The local particle density $\rho$ and hydrodynamic velocity $\mathbf{u}$ remains unchanged in the progress of local chemical reaction, i.e.,
\begin{equation}
\frac{d \rho      }{d t} |_{R(\lambda)}=0 , \label{M_chemical_rho}
\end{equation}
\begin{equation}
\frac{d \mathbf{u}}{d t} |_{R(\lambda)}=0 . \label{M_chemical_u}
\end{equation}
The temperature increases due to the chemical energy released.

4. It is an irreversible reaction, the process of which is described by an empirical or semi-empirical formulas, i.e.,
\begin{equation}
\frac{d\lambda}{dt}=R(\lambda),
\label{lambda_formula}
\end{equation}
where $\lambda$ denotes the progress rate of reaction.

In fact, the chemical term $M$ in Eq.\ref{Boltzmann_chemical} refers to the change of distribution function $f$ due to the local chemical reaction. Specially,
\begin{equation}
M=\frac{df}{dt}|_{R(\lambda)}. \label{M_chemical_1}
\end{equation}
If the physical system has little departure from equilibrium state, $f\approx f^{eq}$, Eq.\ref{M_chemical_1} gives
\begin{equation}
M=\frac{df^{eq}}{dt}|_{R(\lambda)}. \label{M_chemical_2}
\end{equation}
With $f^{eq}=f^{eq}(\rho,\mathbf{u},T)$ and $\frac{d}{dt}=\frac{\partial }{\partial t}+\mathbf{u}\cdot \nabla$, Eq.\ref{M_chemical_2} gives
\begin{equation}
M=\frac{\partial f^{eq}}{\partial \rho}       \frac{d \rho      }{d t} |_{R(\lambda)}
 +\frac{\partial f^{eq}}{\partial \mathbf{u}} \frac{d \mathbf{u}}{d t} |_{R(\lambda)}
 +\frac{\partial f^{eq}}{\partial T   }       \frac{d T         }{d t} |_{R(\lambda)}.
\label{M_chemical_3}
\end{equation}
Substituting Eqs.\ref{M_chemical_rho} and \ref{M_chemical_u} into \ref{M_chemical_3} gives
\begin{equation}
M=\frac{\partial f^{eq}}{\partial T}\frac{d T}{d t}|_{R(\lambda)}.
\label{M_chemical_4}
\end{equation}
With the equilibrium distribution function
\begin{equation}
f^{eq}=\rho (\frac{1}{2 \pi k T})^{D/2} Exp[-\frac{(\mathbf{v}-\mathbf{u})^2}{2kT}],
\end{equation}
for a $D$-dimensional physical system where the particle mass is $m=1$, we get
\begin{equation}
\frac{\partial f^{eq}}{\partial T}=\frac{-D T+(\mathbf{v}-\mathbf{u})^2}{2T^2} f^{eq}. \label{M_chemical_feq}
\end{equation}
With the relation $T=\frac{2E}{D \rho}$ between temperature $T$ and internal energy per volume $E$, we get
\begin{equation}
\frac{d T}{d t}|_{R(\lambda)}=\frac{2}{D \rho}\frac{d E}{d t}|_{R(\lambda)}
=\frac{2Q}{D}\frac{d \lambda}{d t}. \label{M_chemical_T1}
\end{equation}
where $Q$ is the amount of heat released by the chemical reactant per unit mass. Equations \ref{lambda_formula} and \ref{M_chemical_T1} give
\begin{equation}
\frac{d T}{d t}=\frac{2Q}{D} R(\lambda). \label{M_chemical_T2}
\end{equation}
Substituting Eqs.\ref{M_chemical_feq}, \ref{M_chemical_T2} into \ref{M_chemical_4}, we get
\begin{equation}
M=f^{eq} \frac{-D T+(\mathbf{v}-\mathbf{u})^2}{2T^2} \frac{2Q}{D} R(\lambda).
\label{M_continuous}
\end{equation}
It is clear that Eq.\ref{M_continuous} satisfies the following relations
\begin{eqnarray}
\int Md\mathbf{v} &=&\frac{d\rho }{dt}|_{R(\lambda )}=0 \\
\int M\mathbf{v}d\mathbf{v} &=&\frac{d\rho \mathbf{u}}{dt}|_{R(\lambda )}=0
\\
\frac{1}{2}\int M\mathbf{v}^{2}d\mathbf{v} &=&\frac{dE}{dt}|_{R(\lambda
)}=\rho QR(\lambda )
\end{eqnarray}

\subsection{Discrete velocity model}

In a polar coordinate system, the LB equation corresponding to Eq.\ref{Boltzmann_chemical} reads,
\begin{equation}
\frac{\partial f_{ki}}{\partial t}+v_{kir}\frac{\partial f_{ki}}{\partial r}+ \frac{1}{r}v_{ki\theta }\frac{\partial f_{ki}}{\partial \theta }
=-\frac{1} {\tau }( f_{ki}-f_{ki}^{eq})+M_{ki} ,
\label{PC_BGK}
\end{equation}
\begin{equation}
M_{ki}=f_{ki}^{eq} \frac{-D T+(\mathbf{v_{ki}}-\mathbf{u})^2}{2T^2} \frac{2Q}{D} R(\lambda),
\end{equation}
where $r$ ($\theta$) is the radial (azimuthal) coordinate; $f_{ki}$ ($f_{ki}^{eq}$) is the discrete (equilibrium) distribution function; $v_{kir}$ ($v_{ki\theta }$) is the radial (azimuthal) component of the discrete velocity $\mathbf{v}_{ki}$ as below \cite{Watari2011,Watari2003},
\begin{equation}
\begin{array}{c}
\mathbf{v}_{ki}=v_{kir}\mathbf{e}_{r}+v_{kir}\mathbf{e}_{\theta },
\ v_{kir}=v_{k}\cos(i \pi/4-\theta),
\ v_{ki\theta }=v_{k}\sin(i \pi/4-\theta),
\end{array}
\end{equation}
with unit vectors $\mathbf{e}_{r}$ and $\mathbf{e}_{\theta }$. The subscript $k$($=0,1,2,3,4$) indicates the $k$-th\ group of the particle velocities with speed $v_{k}$. One speed is $v_{0}=0$, and each of the other group has $8$ components, i.e. $i=0,1,\cdots$,7. In this work we choose $v_{1}=1.5$, $v_{2}=3.5$, $v_{3}=7.5$, $v_{4}=12.5$.

In terms of local particle density $\rho$ ($=\sum_{ki}f_{ki}$), hydrodynamic velocity $\mathbf{u}$ ($=\sum_{ki}f_{ki}\mathbf{v}_{ki}/\rho$) and temperature $T$ ($=\sum_{ki}\frac{1}{2}f_{ki}(\mathbf{v}_{ki}-\mathbf{u})\cdot(\mathbf{v}_{ki}-\mathbf{u})/\rho$), we get
\begin{eqnarray}
f_{ki}^{eq} &=&\rho F_{k}[(1-\frac{u^{2}}{2T}+\frac{u^{4}}{8T^{2}})+\frac{v_{ki\varepsilon }u_{\varepsilon }}{T}(1-\frac{u^{2}}{2T})+\frac{v_{ki\varepsilon }v_{ki\pi }u_{\varepsilon }u_{\pi }}{2T^{2}}(1-\frac{u^{2}}{2T})  \nonumber\\
&&+\frac{v_{ki\varepsilon }v_{ki\pi }v_{ki\vartheta }u_{\varepsilon }u_{\pi}u_{\vartheta}}{6T^{3}}+\frac{v_{ki\varepsilon }v_{ki\pi }v_{ki\vartheta}v_{ki\xi }u_{\varepsilon }u_{\pi}u_{\vartheta }u_{\xi }}{24T^{4}}]
\label{feq}
\end{eqnarray}
with weighting coefficients
\begin{eqnarray}
F_{k} &=&\frac{1}{v_{k}^{2}(v_{k}^{2}-v_{k+1}^{2})(v_{k}^{2}-v_{k+2}^{2})(v_{k}^{2}-v_{k+3}^{2})}
[48T^{4}-6(v_{k+1}^{2}+v_{k+2}^{2}+v_{k+3}^{2})T^{3} \nonumber\\
&&+(v_{k+1}^{2}v_{k+2}^{2}+v_{k+2}^{2}v_{k+3}^{2}+v_{k+3}^{2}v_{k+1}^{2})T^{2}+\frac{1}{4}v_{k+1}^{2}v_{k+2}^{2}v_{k+3}^{2}T], \nonumber\\
F_{0} &=&1-8(F_{1}+F_{2}+F_{3}+F_{4}) \nonumber,
\end{eqnarray}
where the subscript $\{k+l\}$ equals to $(k+l-4)$ if $(k+l)>4$.

Via the Chapman-Enskog expansion, it is easy to find that Eq.\ref{PC_BGK} could recovery the following Navier-Stokes equations
\begin{equation}
\frac{\partial \rho }{\partial t}+\nabla \cdot (\rho \mathbf{u})=0\text{,}
\label{NS_1}
\end{equation}
\begin{equation}
\frac{\partial (\rho \mathbf{u})}{\partial t}+\nabla \cdot (P\mathbf{I}
+\rho \mathbf{uu})+\nabla \cdot \lbrack \mu (\nabla \cdot \mathbf{u})\mathbf{I}
-\mu (\nabla \mathbf{u})^{T}-\mu \nabla \mathbf{u}]=0 \text{,}
\label{NS_2}
\end{equation}
\begin{gather}
\frac{\partial }{\partial t}(\rho E+\frac{1}{2}\rho u^{2})+\nabla \cdot
\lbrack \rho \mathbf{u}(E+\frac{1}{2}u^{2}+\frac{P}{\rho })]  \notag \\
-\nabla \cdot \lbrack \kappa ^{^{\prime }}\nabla E+\mu \mathbf{u}\cdot
(\nabla \mathbf{u})-\mu \mathbf{u}(\nabla \cdot \mathbf{u})+\frac{1}{2}\mu
\nabla u^{2}]=\rho QR(\lambda ) \text{,}
\label{NS_3}
\end{gather}
in the hydrodynamic limit, where $\mu$(=$P\tau$) and $\kappa ^{^{\prime }}$($=2P\tau$) are viscosity and heat conductivity, respectively.

\subsection{LB evolution equation}

The evolution equation, with first-order accuracy, used for Eq.\ref{PC_BGK}, reads
\begin{equation}
f_{ki}^{t+\Delta t}=f_{ki}^{eq}+(f_{ki}^{t}-f_{ki}^{eq})\exp (-\Delta t/\tau)
+f_{ki,r}^{*}+f_{ki,\theta}^{*}+M_{ki} \Delta t,
\label{LB_evolution_equation}
\end{equation}
and
\begin{equation}
f_{ki,r}^{*} =\left\{
\begin{array}{ccc}
C_{r}[f_{ki}(i_r,i_\theta )-f_{ki}(i_r-1,i_\theta )] & for & C_{r}\geq 0, \\
C_{r}[f_{ki}(i_r+1,i_\theta )-f_{ki}(i_r,i_\theta )] & for & C_{r}<0,
\end{array}
\right.
\end{equation}
\begin{equation}
f_{ki,\theta }^{*} =\left\{
\begin{array}{ccc}
C_{\theta }[f_{ki}(i_r,i_\theta )-f_{ki}(i_r,i_\theta -1)] & for & C_{\theta }\geq 0, \\
C_{\theta }[f_{ki}(i_r,i_\theta +1)-f_{ki}(i_r,i_\theta )] & for & C_{\theta }<0.
\end{array}
\right.
\end{equation}
with Courant-numbers $C_{r}$($=v_{kir}\frac{\Delta t}{\Delta r}$) and $C_{\theta}$($=\frac{1}{r}v_{ki\theta}\frac{\Delta t}{\Delta \theta}$).

\subsection{Boundary conditions}

\begin{figure}
\begin{center}
\includegraphics[bbllx=25pt,bblly=0pt,bburx=585pt,bbury=316pt,width=0.75\textwidth]{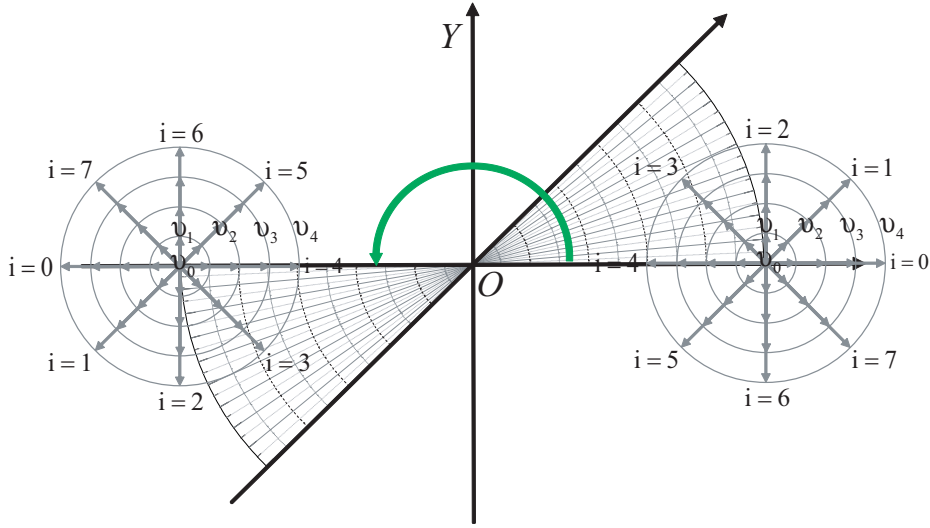}
\end{center}
\caption{Rotation of the distribution functions from the first to the fifth sector of physical domain in a disc divided into $8$ sections.}
\label{Fig01}
\end{figure}
The physical domain under consideration is in a sector which is only $1/8$ of an annular or circular area. The azimuthal boundaries are treated with periodic boundary conditions \cite{XuLin2014}. For annular area with radii $0<R_{1}<R_{2}$, inflow/outflow conditions are imposed at radial boundaries \cite{XuLin2014}. For circular area with radius $R$, its outer radial boundary is treated in the same way. Specially, around the center, the inner boundary is treated as,
\begin{equation}
f(i_r,i_\theta ,k,i)=f(1-i_r,i_\theta,k,mod(i+4,8)) ,
\end{equation}
with $i_r=-1, 0$ for the nodes added to the computational domain and the function $mod(a,b)$ means the remainder of $a$ divided by $b$. Figure \ref{Fig01} shows the relation between the distribution functions in the first and the fifth sector of physical domain in a periodic circular area by rotation.

\subsection{Non-equilibrium characteristics}

LB model naturally inherits the function of Boltzmann equation describing non-equilibrium system. The departure of the system from local thermodynamic equilibrium state can be measured by the high-order moments of $f_{ki}$. As given in \cite{XuYan2013,XuLin2014}, the central moments $\mathbf{M}_{m}^*$ are defined as:
\begin{eqnarray}
&&\left\{
\begin{array}{l}
\mathbf{M}_{2}^*(f_{ki})=\sum_{ki}f_{ki}\mathbf{v}_{ki}^*\mathbf{v}_{ki}^* \\
\mathbf{M}_{3}^*(f_{ki})=\sum_{ki}f_{ki}\mathbf{v}_{ki}^*\mathbf{v}_{ki}^*\mathbf{v}_{ki}^* \\
\mathbf{M}_{3,1}^*(f_{ki})=\sum_{ki}\frac{1}{2}f_{ki}\mathbf{v}_{ki}^*\cdot\mathbf{v}_{ki}^*\mathbf{v}_{ki}^* \\
\mathbf{M}_{4,2}^*(f_{ki})=\sum_{ki}\frac{1}{2}f_{ki}\mathbf{v}_{ki}^*\cdot\mathbf{v}_{ki}^*\mathbf{v}_{ki}^*\mathbf{v}_{ki}^*
\end{array}
\right.
\label{M_absolute}
\end{eqnarray}
where $\mathbf{v}_{ki}^*=\mathbf{v}_{ki}-\mathbf{u}$. The manifestations of non-equilibrium are defined as:
\begin{eqnarray}
\mathbf{\Delta }_{m}^{* } &=&\mathbf{M}_{m}^{* }(f_{ki})-\mathbf{M}_{m}^{* }(f_{ki}^{eq})  .
\label{D_absolute}
\end{eqnarray}
In theory, $\mathbf{M}_{3}^{* }(f_{ki}^{eq})=0$, $\mathbf{M}_{3}^{* }(f_{ki})=\mathbf{\Delta }_{3}^{* }$, $\mathbf{M}_{3,1}^{*}(f_{ki}^{eq})=0$, $\mathbf{M}_{3,1}^{* }(f_{ki})=\mathbf{\Delta }_{3,1}^{* }$, $\mathbf{M}_{3,1,r}^*(f_{ki})=\frac{1}{2} (\mathbf{M}^*_{3,rrr}(f_{ki})+\mathbf{M}^*_{3,r\theta\theta}(f_{ki}))$, $\mathbf{M}_{3,1,\theta}^*(f_{ki})=\frac{1}{2} (\mathbf{M}^*_{3,rr\theta}(f_{ki})+\mathbf{M}^*_{3,\theta\theta\theta}(f_{ki}))$, $\mathbf{\Delta}_{3,1,r}^*(f_{ki})=\frac{1}{2} (\mathbf{\Delta}^*_{3,rrr}(f_{ki})+\mathbf{\Delta}^*_{3,r\theta\theta}(f_{ki}))$, $\mathbf{\Delta}_{3,1,\theta}^*(f_{ki})=\frac{1}{2} (\mathbf{\Delta}^*_{3,rr\theta}(f_{ki})+\mathbf{\Delta}^*_{3,\theta\theta\theta}(f_{ki}))$.

\section{Detonation}

Detonation is in a complex process with mutual influence between fluid dynamics and chemical reaction kinetics. The detonation front propagates into unburnt gas at a velocity higher than the speed of sound in front of the wave \cite{Bjerketvedt1997}. Physical quantities at two sides of detonation front satisfy Hugoniot relations \cite{Fickett1979}.

\subsection{Chemical reaction}

To describe the chemical process of detonation, we choose Cochran's rate function presented by Cochran and Chan \cite{Cochran1979},
\begin{equation}
R(\lambda)=\omega_{1}P^{m}(1-\lambda )+\omega_{2}P^{n}\lambda(1-\lambda ) ,
\label{Cochran}
\end{equation}
where $\lambda $($=\rho _{p}/\rho $) is the mass fraction of reacted reactant, and $\rho _{p}$ is the density of reacted reactant. The right side of Eq.\ref{Cochran} is composed of a hot formation term and a growth term. $P^{m}$ and $P^{n}$ describe the dependence on the local pressure and $\omega_{1}$, $\omega_{2}$, $m$ and $n$ are adjustable parameters. Furthermore, $T>T_{th}$ is a necessary condition for chemical reaction, with the ignition temperature $T_{th}$. In this work, we choose $m=n=1$, $T_{th}=1.1$.

Via introducing the symbol, $a=\omega _{1}P^{m}$, $b=\omega _{2}P^{n}$, $\lambda=\lambda_{i_r}=\lambda_{i_\theta}=\lambda(i_r,i_\theta,t)$, the evolution of Eq.\ref{Cochran} with first-order accuracy reads
\begin{equation}
\lambda ^{t+\Delta t}=\frac{(a+b\lambda )e^{(a+b)\Delta t}-a(1-\lambda )}
{(a+b\lambda )e^{(a+b)\Delta t}+a(1-\lambda )}+\lambda_{i_r}^{*}+\lambda _{i_\theta }^{*} ,
\label{Eqlambda}
\end{equation}
and
\begin{equation}
\lambda _{i_r}^{* }=\left\{
\begin{array}{ccc}
-\frac{u_{r}(\lambda _{i_r}-\lambda _{i_r-1})}{\Delta r}\Delta t & for & u_{r}\geq 0, \\
-\frac{u_{r}(\lambda _{i_r+1}-\lambda _{i_r})}{\Delta r}\Delta t & for & u_{r}<0,
\end{array}
\right.
\end{equation}
\begin{equation}
\lambda _{i_\theta }^{* }=\left\{
\begin{array}{ccc}
-\frac{u_{\theta }(\lambda _{i_\theta}-\lambda _{i_\theta -1})}{r\Delta \theta}\Delta t & for & u_{\theta }\geq 0, \\
-\frac{u_{\theta }(\lambda _{i_\theta +1}-\lambda _{i_\theta})}{r\Delta \theta}\Delta t & for & u_{\theta }<0.
\end{array}
\right.
\end{equation}
In the evolution of $\lambda$, the velocity $\mathbf{u}$ and the pressure $P$ are calculated from $f_{ki}$. In this way, the chemical reaction has coupled naturally with the flow behaviors.

It is worth pointing out that the Cochran's rate function used in this work is similar to, but different from, the Lee-Tarver model used in the work \cite{XuYan2013}. The parameters $a$ and $b$ in Eq.\ref{Eqlambda} depend on the pressure in the former work, while they are given fixed values in the latter work where the pressure plays no role in the chemical process. Consequently, the extinction phenomenon can be investigated in this work and can not be simulated in the latter work.

\subsection{Simulation of steady detonation}

\subsubsection{Validation and verification}

In this section, a steady detonation is simulated to demonstrate the validity of the new model. The initial physical quantities are as:
\begin{equation}
\left\{
\begin{array}{l}
(\rho ,T,u_{r},u_{\theta },\lambda )_{i}=(1.35826,2.59709,0.81650,0,1)\\
(\rho ,T,u_{r},u_{\theta },\lambda )_{o}=(1,1,0,0,0)
\end{array}
\right.
\label{V_and_V}
\end{equation}
which satisfy the Hugoniot relations for detonation wave. Here the suffixes $i$ and $o$ index two parts, $1000 \le r \le 1000.01$ and $1000.01<r\le 1000.1$, in an annular area, respectively. The inner radius is given large enough, so that the curvature becomes negligible and the polar coordinates revert locally to Cartesian coordinates. With this condition, the simulation results can be compared to the analytic solutions of the 1-dimensional steady detonation wave. Other parameters are $\tau=2\times 10^{-4}$, $\Delta t=2.5\times 10^{-7}$, $N_{r}\times N_{\theta }=20000\times 1$, $\omega _{1}=10$, $\omega_{2}=1000$.

\begin{figure}
\begin{center}
\includegraphics[bbllx=0pt,bblly=303pt,bburx=595pt,bbury=611pt,width=1.0\textwidth]{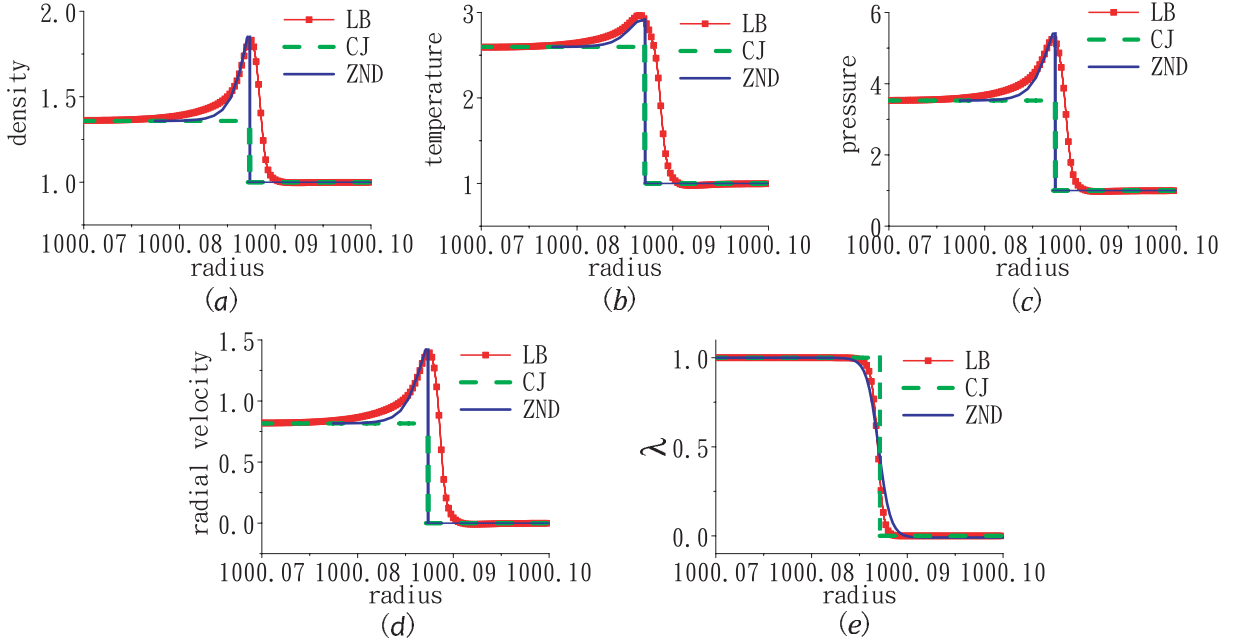}
\end{center}
\caption{physical quantities of steady detonation wave at time $t=0.025$, with radial range $1000.07\le r\le 1000.1$: (a) $\rho$, (b) $T$, (c) $P$, (d) $u_r$, (e) $\lambda$.}
\label{Fig02}
\end{figure}
Figure \ref{Fig02} gives LB simulation results, CJ results \cite{Fickett1979,Chapman1899,Jouguet1905} and ZND results \cite{Fickett1979,Zeldovich1940,Neumann1942,Doering1943} of physical quantities ($\rho$, $T$, $P$, $u_r$, $\lambda$) at time $t=0.025$, with radial range $1000.07\le r\le 1000.1$, respectively. The solid lines with squares are for LB simulation results, the dashed lines are for analytic solutions of CJ theory, and the solid lines are for analytic solutions of ZND theory. The simulation physical quantities after detonation wave are $(\rho,T,u_{r},u_{\theta},\lambda)=(1.36163,2.59366,0.819670,0,1)$. Comparing them with
CJ results gives the relative differences $0.2\%$, $0.1\%$, $0.3\%$, $0\%$ and $0\%$, respectively. Panels (a)-(e) show that the LB simulation results have a satisfying agreement with the ZND results in the area behind von Neumann peak. There are few differences between them. Physically, the analytic solutions of ZND theory here ignore the viscosity and heat conduction, and the von Neumann peak is simply treated as a strong discontinuity. Furthermore, the relative difference is $1.8\%$ between the simulation detonation velocity $D=3.152$ and the analytic solution $D=3.09557$. In sum, the current LB model works for detonation phenomenon.

\subsubsection{Nonequilibriums in steady detonation wave}

\begin{figure}
\begin{center}
\includegraphics[bbllx=22pt,bblly=234pt,bburx=563pt,bbury=474pt,width=0.75\textwidth]{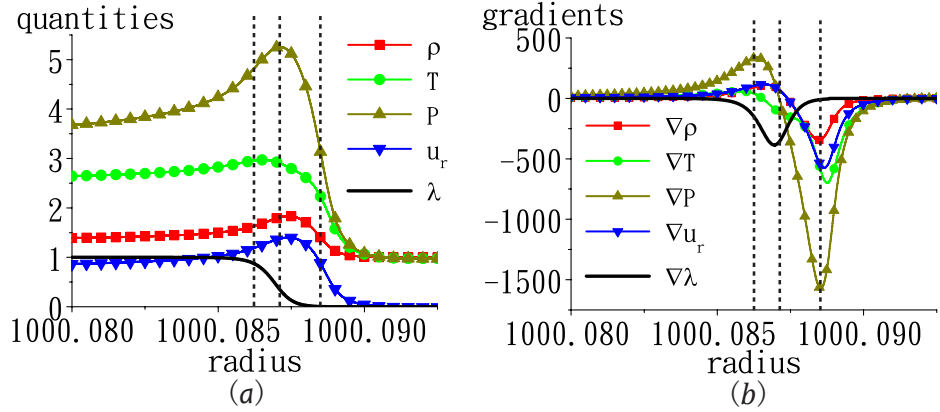}
\end{center}
\caption{The profile of steady detonation wave in an annular area with radii $R_{1}=1000$ and $R_{2}=1000.1$ at time $t=0.025$: (a) physical quantities, (b) gradients. From left to right, three vertical lines are shown to guide the eyes for the rarefaction area, the maximum value of pressure, the pre-shocked area, respectively. }
\label{Fig03}
\end{figure}
Figure \ref{Fig03} gives the physical quantities and their gradients versus radius at time $t=0.025$, with radial range $1000.08\le r\le 1000.0925$. Three vertical lines are shown, from left to right, to guide the eyes for the rarefaction area, the von-Neumann peak and the pre-shocked area, respectively. Panel (a) shows that the maximum values of density, temperature, pressure, velocity do not coincide. The radial positions of their maximum values are $R_{\rho }=1000.08740$, $R_{T}=1000.08654$, $R_{P}=1000.08712$, $R_{u}=1000.08739$. Panel (b) shows that the largest absolute values of $\nabla \rho$, $\nabla T$, $\nabla P$, $\nabla u$ are located at the pre-shocked area, their second largest values are at the rarefaction area, and their vales are close to zero at the von-Neumann peak.

\begin{figure}
\begin{center}
\includegraphics[bbllx=6pt,bblly=278pt,bburx=573pt,bbury=641pt,width=0.99\textwidth]{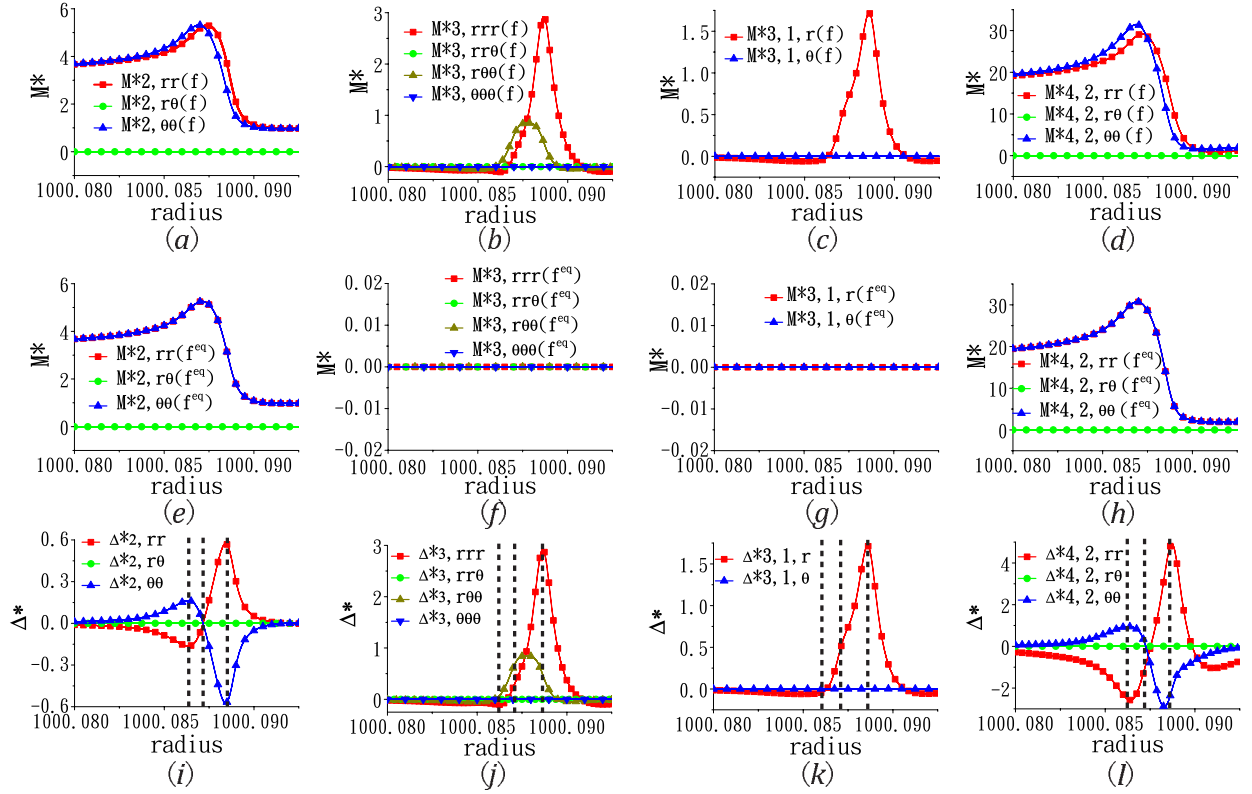}
\end{center}
\caption{The simulation results of $\mathbf{M}_{m}^*(f_{ki})$, $\mathbf{M}_{m}^*(f_{ki}^{eq})$ and $\mathbf{\Delta }_{m}^{*}$ in the same case as Fig.\ref{Fig03}.}
\label{Fig04}
\end{figure}
Figure \ref{Fig04} shows the central moments and their non-equilibrium manifestations in the case corresponding to Fig.\ref{Fig03}. The simulation results of $\mathbf{M}_{2}^{*}(f_{ki})$, $\mathbf{M}_{3}^{*}(f_{ki})$, $\mathbf{M}_{3,1}^{*}(f_{ki})$, $\mathbf{M}_{4,2}^{*}(f_{ki})$, $\mathbf{M}_{2}^{*}(f_{ki}^{eq})$, $\mathbf{M}_{3}^{*}(f_{ki}^{eq})$, $\mathbf{M}_{3,1}^{*}(f_{ki}^{eq})$, $\mathbf{M}_{4,2}^{*}(f_{ki}^{eq})$, $\mathbf{\Delta }_{2}^{*}$, $\mathbf{\Delta }_{3}^{*}$, $\mathbf{\Delta}_{3,1}^{*}$, $\mathbf{\Delta }_{4,2}^{*}$ are shown in Figs.\ref{Fig04} (a)-(l), respectively. The vertical lines in Figs.\ref{Fig04} (i)-(l) coincide with the ones in Fig.\ref{Fig03}. It's easy to get from Fig.\ref{Fig04} that, the non-equilibrium system is mainly around the von-Neumann peak. The departure of the system from its equilibrium around the rightmost line is opposite the one around the leftmost line. Physically, the former is under shock effect, whereas the latter under rarefaction effect. Furthermore, both $\Delta _{2,rr}^{*}$ and $\Delta_{2,\theta \theta }^{*}$ are close to zero at the von-Neumann peak, i.e., the internal kinetic energies in different degrees of the freedom are equal to each other at the von-Neumann peak. Comparing Fig.\ref{Fig03} (b) with Fig.\ref{Fig04} (i) gives that the internal kinetic energies in different degrees of freedom show the maximum difference at the inflexion point where the pressure has the largest spatial derivative. Figure \ref{Fig04} (a) shows that the internal energy in the freedom of $r$ and that in the freedom of $\theta$ do not coincide due to the fluid viscosity. The former travels faster than the latter around the detonation front.

\begin{figure}
\begin{center}
\includegraphics[bbllx=0pt,bblly=0pt,bburx=369pt,bbury=152pt,width=0.6\textwidth]{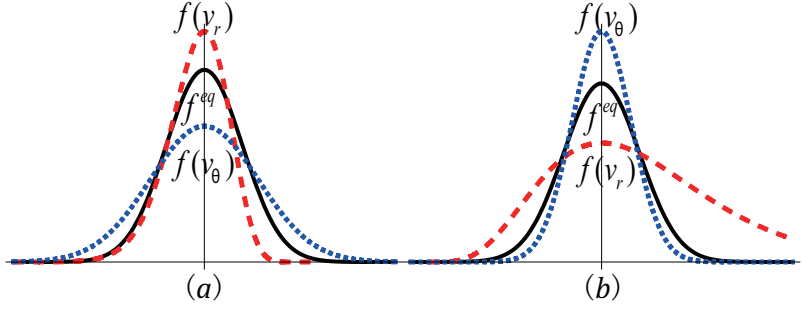}
\end{center}
\caption{The sketch of the Maxwellian and actual distribution functions versus velocity $v_{r}$ and $v_{\protect\theta }$, respectively. (a) the distribution functions at the leftmost line, (b) the distribution functions at the rightmost line. The long-dashed line is for distribution function $f(v_{r})$, the shot-dashed one is for distribution function $f(v_{\theta})$, and the solid line is for $f^{eq}$ .}
\label{Fig05}
\end{figure}
Around the leftmost line in Fig.\ref{Fig04} (i), $\mathbf{\Delta}_{2,rr}^{*}$ shows a negative peak and $\mathbf{\Delta}_{2,\theta \theta }^{*}$ shows a positive peak with the same amplitude, which implies that the distribution function $f(v_{r})$ is \textquotedblleft thinner\textquotedblright and \textquotedblleft higher\textquotedblright than the Maxwellian $f^{eq}$, and $f(v_{\theta })$ is \textquotedblleft fatter\textquotedblright and \textquotedblleft lower\textquotedblright. The simulation results of $\mathbf{\Delta }_{3}^{* }$ in Fig.\ref{Fig04}(j) and $\mathbf{\Delta }_{3,1}^{* }$ in Fig.\ref{Fig04}(k) indicate that $f(v_{\theta })$ is symmetric, and the $f(v_{r})$ is asymmetric. The portion of $f(v_{r})$ for $v_{r}>0$ is \textquotedblleft thinner\textquotedblright than that for $v_{r}<0$. Figure \ref{Fig05} (a) shows the sketch of the actual distribution functions, $f(v_{r})$, $f(v_{\theta })$ and the Maxwellian $f^{eq}$. For the rightmost line in Fig.\ref{Fig04}, similarly, Fig.\ref{Fig05}(b) shows the sketch of the actual distribution functions, $f(v_{r})$, $f(v_{\theta })$ and the Maxwellian $f^{eq}$. It can be found that $f(v_{r})$ is \textquotedblleft fatter\textquotedblright and \textquotedblleft lower\textquotedblright than the Maxwellian $f^{eq}$, while $f(v_{\theta })$ is \textquotedblleft thinner\textquotedblright and \textquotedblleft higher\textquotedblright. $f(v_{\theta })$ is symmetric, while $f(v_{r})$ is asymmetric. The portion of $f(v_{r})$\ for $v_{r}>0$ is \textquotedblleft fatter\textquotedblright\ and the portion of $f(v_{r})$ for $v_{r}<0$ is \textquotedblleft thinner\textquotedblright .

Moreover, the simulation result $\mathbf{\Delta }_{2,r\theta}^{*}=0$ in Fig.\ref{Fig04}(i) indicate that the contours of the actual distribution function in velocity space ($v_{r}$,$v_{\theta }$) is symmetric about the $v_{r}$-axis or/and $v_{\theta }$-axis. The above analysis suggests that $v_{r}$-axis is the symmetric axis. Combining Figs.\ref{Fig05} and the sketch of contours of the actual distribution function gives the sketch of the actual distribution function in velocity space ($v_{r}$,$v_{\theta }$) \cite{XuLin2014}.

\subsection{Simulation of implosion}

For the case of implosion, the initial physical quantities are:
\begin{equation}
\left\{
\begin{array}{l}
(\rho ,T,u_{r},u_{\theta },\lambda )_{i}=(1,1,0,0,0) \\
(\rho ,T,u_{r},u_{\theta },\lambda )_{o}=(1.5,1.55556,-0.666667,0,1) ,
\end{array}
\right.
\end{equation}
where the suffixes $i$ and $o$ index areas $0\le r\le 0.098$ and $0.098<r\le 0.1$, respectively. Other parameters are $\tau =2\times 10^{-4}$, $\Delta t=2.5 \times 10^{-6}$, $N_{r}\times N_{\theta}=2000\times 1$, $\omega_{1}=1$, $\omega_{2}=50$.
\begin{figure}
\begin{center}
\includegraphics[bbllx=0pt,bblly=0pt,bburx=575pt,bbury=382pt,width=0.99\textwidth]{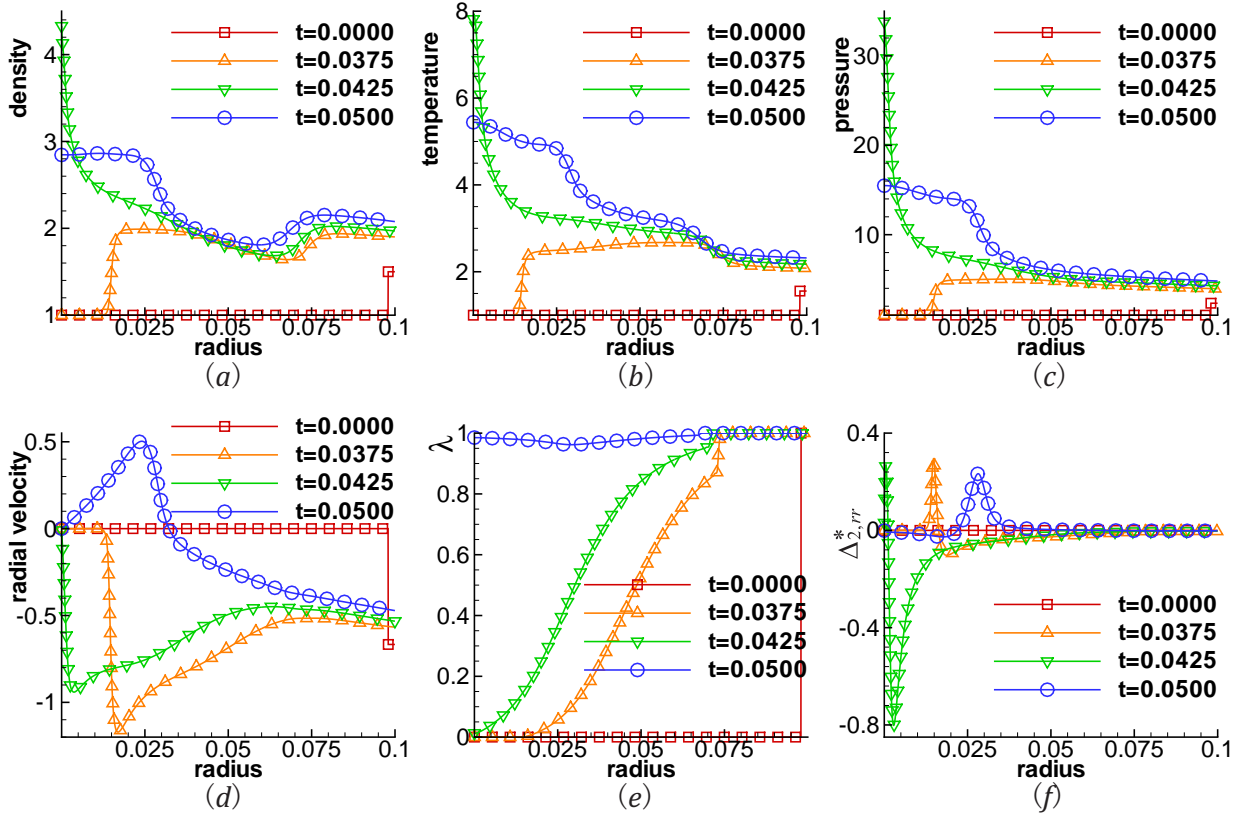}
\end{center}
\caption{Physical quantities versus radius in implosion process at times, $t=0.0000$, $0.0375$, $0.0425$ and $0.0500$, respectively: (a) $\rho$; (b) $T$; (c) $P$; (d) $u_{r}$; (e) $\lambda$; (f) $\Delta_{2,rr}^{*}$. }
\label{Fig06}
\end{figure}

Figure \ref{Fig06} shows physical quantities along radius in implosion process: (a) density; (b) temperature; (c) pressure; (d) radial velocity; (e) the parameter for chemical reaction process; (f) $\Delta_{2,rr}^{*}$. There are two stages in implosion process. In the former stage, the detonation travels inwards, the material behind the detonation front moves inwards, and the density, temperature and pressure behind detonation wave increase continuously due to the disc geometric effect. When the detonation wave reaches the center, the density, temperature and pressure get their maximum values. Meanwhile, the velocity reduces to zero gradually and then point outwards. In the latter stage, the detonation wave travels outwards. As the chemical reaction completes, the detonation wave becomes a pure shock wave. The hydrodynamic velocity in front of the shock wave points inwards and that behind the wave points outwards. Consequently, the density, temperature and pressure outside the shock wave still increase continuously, and those inside reduce.

In addition, Fig.\ref{Fig06} (f) shows that the departure of the system from equilibrium increases (reduces) when the detonation or shock wave becomes stronger (weaker). Specially, the value of $\Delta_{2,rr}^{*}$ shows a crest and a trough from the time $t=0.0000$ to $t=0.0425$. The crest results from compression effect ahead of the detonation front, while the trough results from rarefaction effect behind. From $t=0.0425$ to $0.0500$, it is also positive at the crest and negative behind. In fact, $\Delta_{2,rr}^{*}$ is always positive at the shock wave and negative at the rarefaction wave, which can be seen as a criterion to distinguish the two waves. Furthermore, the system at the disc center is always in its thermodynamic equilibrium.

\subsection{Simulation of explosion}

\begin{figure}
\begin{center}
\includegraphics[bbllx=0pt,bblly=0pt,bburx=578pt,bbury=381pt,width=0.99\textwidth]{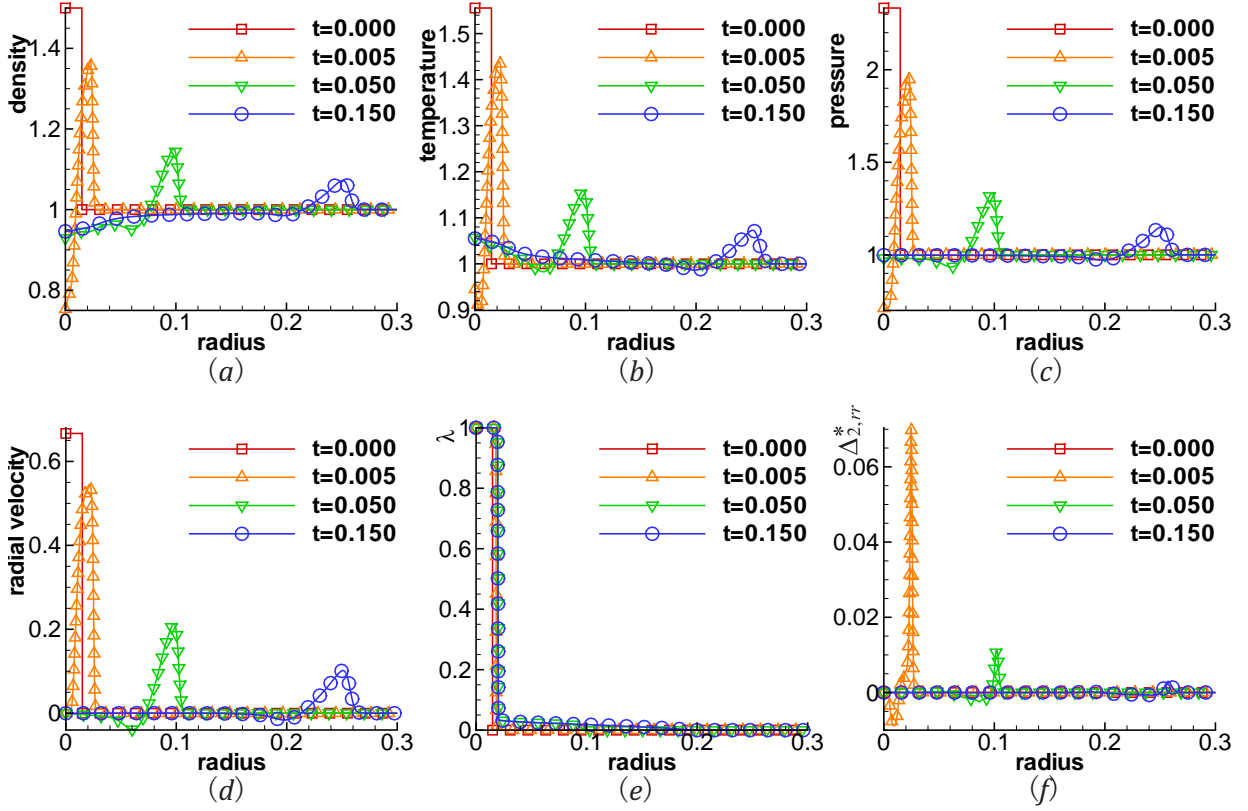}
\end{center}
\caption{For the case $R_{1}=0.015$, physical quantities versus radius in explosion process at times, $t=0.000$, $0.005$, $0.050$ and $0.150$, respectively: (a) $\rho$; (b) $T$; (c) $P$; (d) $u_{r}$; (e) $\lambda$; (f) $\Delta_{2,rr}^{*}$. }
\label{Fig07}
\end{figure}

\begin{figure}
\begin{center}
\includegraphics[bbllx=0pt,bblly=0pt,bburx=580pt,bbury=383pt,width=0.99\textwidth]{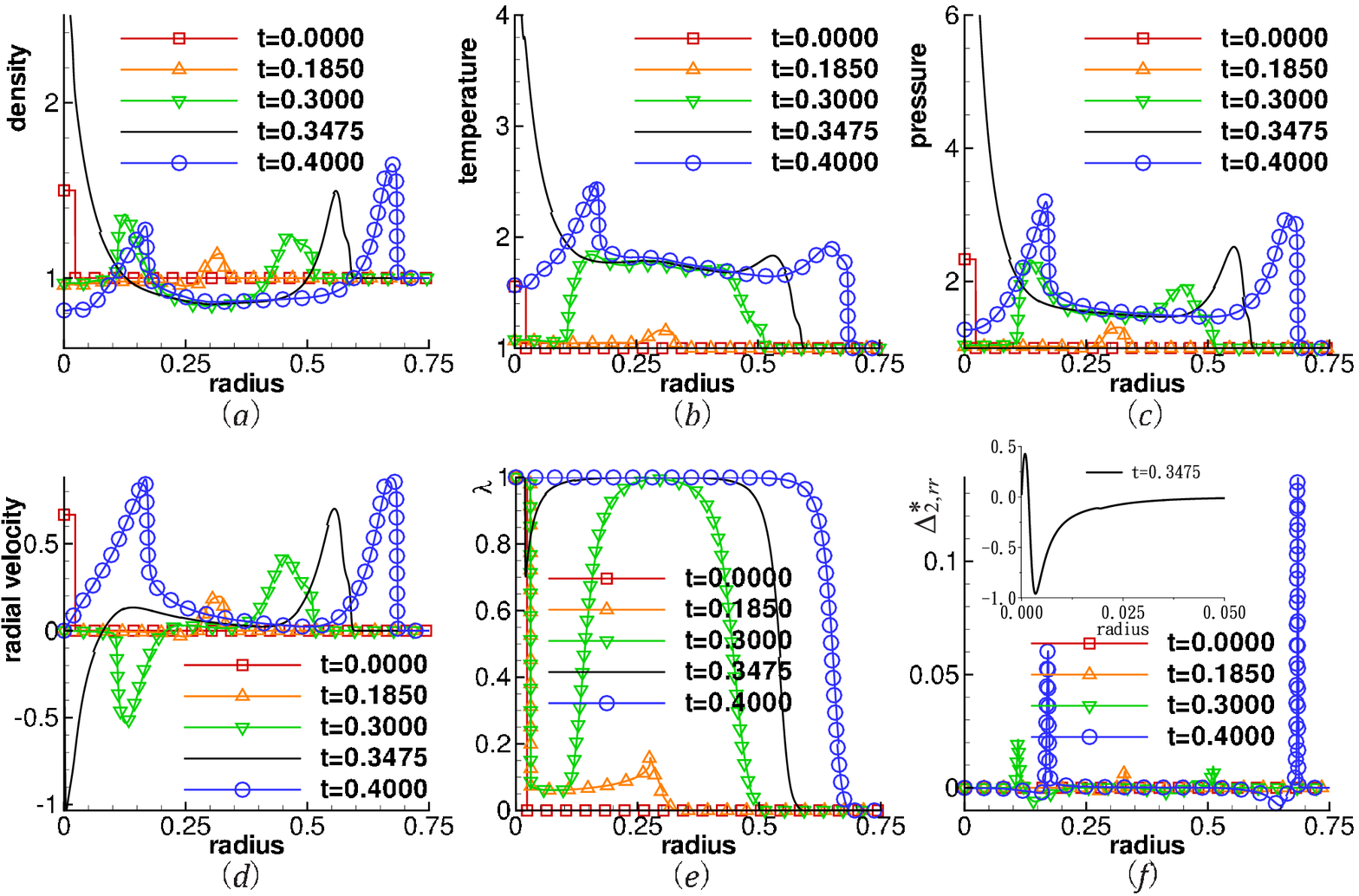}
\end{center}
\caption{For the case $R_{1}=0.023$, physical quantities versus radius in explosion process at times, $t=0.0000$, $0.1850$, $0.3000$, $0.3475$ and $0.4000$, respectively: (a) $\rho$; (b) $T$; (c) $P$; (d) $u_{r}$; (e) $\lambda$; (f) $\Delta_{2,rr}^{*}$. }
\label{Fig08}
\end{figure}

\begin{figure}
\begin{center}
\includegraphics[bbllx=0pt,bblly=0pt,bburx=577pt,bbury=386pt,width=0.99\textwidth]{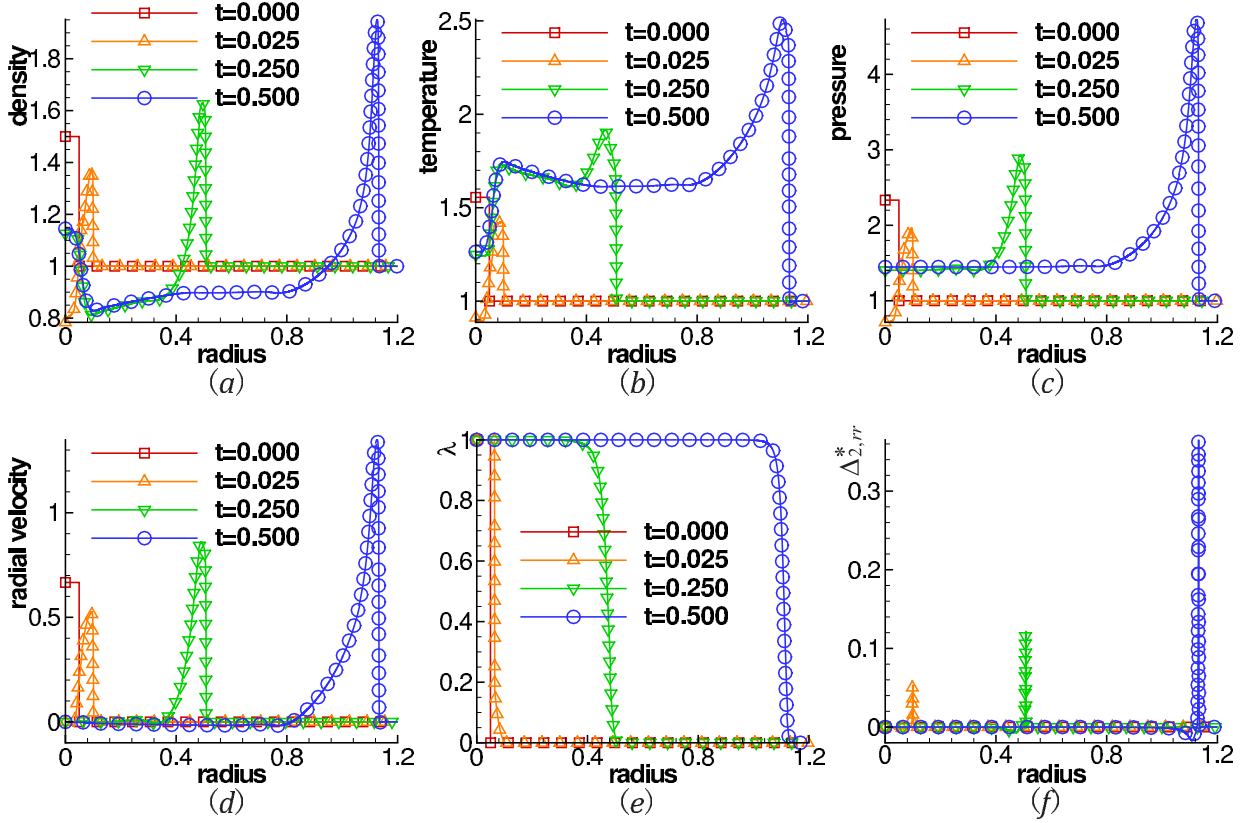}
\end{center}
\caption{For the case $R_{1}=0.050$, physical quantities versus radius in explosion process at times, $t=0.000$, $0.025$, $0.250$ and $0.500$, respectively: (a) $\rho$; (b) $T$; (c) $P$; (d) $u_{r}$; (e) $\lambda$; (f) $\Delta_{2,rr}^{*}$. }
\label{Fig09}
\end{figure}

For the case of explosion, the initial physical quantities are:
\begin{equation}
\left\{
\begin{array}{l}
(\rho ,T,u_{r},u_{\theta },\lambda )_{i}=(1.5,1.55556,0.666667,0,1) \\
(\rho ,T,u_{r},u_{\theta },\lambda )_{o}=(1,1,0,0,0)
\end{array}
\right.
\end{equation}
where the suffixes $i$ and $o$ index areas $0\le r\le R_{1}$ and $R_{1}<r\le R$, respectively. Here $\tau =2\times 10^{-4}$, $\Delta t=2.5\times 10^{-6}$, $\omega _{1}=1$, $\omega _{2}=50$. Figures \ref{Fig07}-\ref{Fig09} show the evolution of physical quantities ($\rho$, $T$, $P$, $u_{r}$, $\lambda$, $\Delta_{2,rr}^{*}$) versus radius. Figure \ref{Fig07} corresponds to parameters $R_{1}=0.015$, $R=0.3$, $N_{r}\times N_{\theta }=6000\times 1$; Fig.\ref{Fig08} corresponds to parameters $R_{1}=0.023$, $R=0.75$, $N_{r}\times N_{\theta }=15000\times 1$; Fig.\ref{Fig09} corresponds to parameters $R_{1}=0.050$, $R=1.2$, $N_{r}\times N_{\theta }=24000\times 1$.

Figure \ref{Fig07} shows the extinction phenomenon. The area of initial reacted reaction is very small, i.e., the initial energy is not enough to trigger the detonation. The energy propagates outward in the form of disturbance wave with amplitude decreasing gradually. The wave dissipates and vanishes finally. It's easy to get from Fig.\ref{Fig07} (e) that $\lambda $ only has a little change at the beginning. The reason is that the initial temperature of reacted reaction is higher than the temperature threshold $T_{th}$ and there is a little chemical reaction at the start. The rate of chemical energy released is smaller than the rate of heat dissipated under the disc geometric effect. Consequently, the energy of disturbance wave reduces gradually, which leads to extinction. Figure (f) shows that the system has only a small departure from equilibrium at the start. The departure reduces gradually and vanishes finally.

Figure \ref{Fig08} shows that (i) A disturbance wave travels outwards from $t=0.0000$ to $0.1850$. Part reactant reacts around the disturbance wave. The released chemical energy is added into the disturbance wave, meanwhile the thermal energy within disturbance wave disperses to its adjacent area. The disturbance wave becomes wider in both radical and azimuthal directions under disc geometric effect, and the maximum values of density, temperature, pressure and velocity reduce. The value of $\Delta_{2,rr}^{*}$ shows a crest at the disturbance wave and a trough behind. (ii) The perturbation wave is transformed into detonation wave from $t=0.1850$ to $0.3000$. The heat release rate of chemical reactant increases sharply. The physical quantities ($\rho$, $T$, $P$, $u_{r}$) increase suddenly. Meanwhile the number of crest (trough) of $\Delta_{2,rr}^{*}$ increases from one to two. (iii) The implosion and explosion waves coexist from $t=0.3000$ to $0.3475$. With amplitude defined as the distance from the crest to trough of $\Delta_{2,rr}^{*}$, it is found that the amplitudes at the implosion and explosion waves increase dramatically, especially the former one increases from $2.78\times10^{-2}$ to $1.39$. For the purpose of a clear view, the plot of $\Delta_{2,rr}^{*}$ at the time $t=0.3475$ is given specially within panel (f). (iv) From $t=0.3475$ to $0.4000$, the implosion wave passes the center of the disc, then travels outwards and changes into a shock wave with the completion of chemical reaction. It is then transformed into a perturbation wave, dissipates and vanishes finally. While the explosion wave propagates outwards and its peak rises further. The amplitude of $\Delta_{2,rr}^{*}$ at the inner outgoing wave reduces, and the one at the explosion wave increases.

Figure \ref{Fig09} shows a direct explosion phenomenon with ignition energy large enough. Panel (a) shows that the density behind the detonation front is lower than the one outside due to the disc geometric effect. Panels (b)-(c) shows higher temperature and pressure inside. Panel (d) shows that the velocity inside is close to zero. Panel (e) shows that chemical reaction steadily proceeds. Panel (f) shows that the amplitude of $\Delta_{2,rr}^{*}$ increases gradually. The detonation wave becomes wider and wider in both radical and azimuthal directions.

In fact, there is competition between the chemical reaction, macroscopic transportation, thermal diffusion and the geometric convergence or divergence in the detonation phenomenon. The chemical reaction increases the temperature while the thermal diffusion decreases the temperature around the detonation wave. If there is enough thermal energy transformed from chemical energy, the detonation proceeds; otherwise, extinguishes. Specially in explosion case, if the geometric divergence effects dominate, extinction will occur; with the the combustion front propagating outwards, geometric divergence makes less effect, the existing part combustion may result in complete combustion, even detonation.

\section{Conclusions and discussions}

A polar coordinate Lattice Boltzmann Kinetic Model (LBKM) for detonation phenomena is presented. Within this novel model, the change of discrete distribution function due to local chemical reaction is given directly in the modified lattice Boltzmann equation, which could recovery the Navier-Skokes equations including chemical reaction via Chapman-Enskog expansion. And the chemical reaction is described by Cochran's rate function. A combined scheme is used to treat with the LB equation and the Cochran's rate equation. Both the temporal evolution of the collision effects in the LB equation and the temporal evolution of chemical reaction in Cochran's rate equation are calculated analytically. Both the convection terms in the LB equation and Cochran's rate equation are treated using the first-order upwind scheme. From the numerical point of view, compared with the LBKM in Ref. \cite{XuLin2014}, the present model has the same accuracy but is simpler. From the physical or chemical point of view, compared with the Lee-Tarver model used in previous work \cite{XuYan2013}, the pressure effects on the reaction rate is taken into account in the Cochran's rate function.

Compared with the previous work in Ref. \cite{XuLin2014}, the inner boundary condition for the disc computational domain is treated more naturally. In Ref. \cite{XuLin2014} the disc computational domain is approximated by a annular domain where the inner radius approaches zero. Consequently, one needs to construct ghost nodes for the inner boundary condition. In this work the center of the disc computational domain is considered as a inner point of the system. For periodic system, no ghost node is needed for the inner boundary condition. Other boundaries are treated with the same method as \cite{XuLin2014}.

The simulation results of physical quantities in the steady detonation process have a satisfying agreement with analytical solutions. Typical implosion and explosion phenomena are simulated. By changing initial ignition energy, we investigate three cases of explosion, including a case with extinction phenomenon. It is interesting to find that the geometric convergence or divergence effect makes the detonation procedure much more complex. The competition between the chemical reaction, the macroscopic transportation, the thermal diffusion and the geometric convergence or divergence determines the ignition process. If there is enough thermal energy transformed from chemical energy, the detonation proceeds; otherwise, extinguishes. Specially in explosion case, if the geometric divergence effects dominate, extinction will occur; with the combustion front propagating outwards, geometric divergence makes less effect, the existing part combustion may result in complete combustion, even detonation.

Moreover, the non-equilibrium behaviors in detonation phenomenon are investigated via the velocity moments of discrete distribution functions. The system at the disc center is always in its thermodynamic equilibrium. The internal kinetic energies in different degrees of freedom around the detonation front do not coincide due to the fluid viscosity. They show the maximum difference at the inflexion point where the pressure has the largest spatial derivative. The influence of shock strength on the reaction rate and the influences of both the shock strength and the reaction rate on the departure amplitude of the system from its local thermodynamic equilibrium are probed. The departure from equilibrium in front of von-Neumann peak results from shock effect, while the one behind the peak results from rarefaction effect. The departure increases when the shock or rarefaction effect increases. Specially, the value of $\Delta_{2,rr}^{*}$ is positive at shock wave and negative at the rarefaction wave, which can be seen a criterion to distinguish the two waves. What's more, the main behaviors of actual distribution functions around the detonation wave are recovered from the numerical results of high-order moments of the discrete distribution function.

Finally, further discussions include the following points.
(i) For the combustion systems where both the reactant and product have more than one components, a multi-distribution-function model is more preferred. Such a work is in progress \cite{XuLin_multi_distribution}.
(ii) The transport properties are relevant to the relaxation time $\tau$.  It should depend on physical quantities, such as density and temperature. Consequently, it should be a function of the space and time. This point should be further investigated in the future.
(iii) In numerical simulations the spatial and temporal steps should be small enough so that the spurious transportation is negligible compared with the physical one.
\section*{Acknowledgements}

The authors thank Prof. Cheng Wang for many helpful discussions. AX and GZ acknowledge support of the Science Foundations of National Key Laboratory of Computational Physics, National Natural Science Foundation of China and the opening project of State Key Laboratory of Explosion Science and Technology (Beijing Institute of Technology) [under Grant No. KFJJ14-1M]. YL and CL acknowledge support of National Natural Science Foundation of China [under Grant No. 11074300], National Basic Research Program of China (under Grant No. 2013CBA01504) and National Science and Technology Major Project of the Ministry of Science and Technology of China (under Grant No.2011ZX05038-001).




\begin{thebibliography}{00}

\bibitem{Fickett1979} W. Fickett and W. C. Davis, \textit{Detonation}, University of California Press, Berkeley, 1979.

\bibitem{Cheng2004} W. Y. Cheng, X. Y. Liu, K. J. Wang, et al, Journal of China Coal Society, 29 (2004) 57 (in Chinese).

\bibitem{Li2010} X. J. Li, B. Q. Lin, Coal Geology \& Exploration, 38 (2010) 7 (in Chinese).

\bibitem{Chapman1899} D. L. Chapman, Philos. Mag, 47 (1899) 90.

\bibitem{Jouguet1905} E. J. Jouguet, Math. Pures Appl, 1 (1905) 347.

\bibitem{Zeldovich1940} Ya. B. Zeldovich, J. Exp. Th. Phys., 10 (1940) 542.

\bibitem{Neumann1942} J. Von Neumann, Theory of Detonation Waves, New York: Macmillan, 1942.

\bibitem{Doering1943} W. Doering, Ann. Phys., 43 (1943) 421.

\bibitem{Bjerketvedt1997} D. Bjerketvedt, J. R. Bakke, K. Van Wingerden, J. Hazard. Mater, 52 (1997) 1.

\bibitem{XunKun2000} Y. Lian, K. Xu, J. Comput. Phys., 163 (2000) 349.

\bibitem{Wang2012} C. Wang, X. Zhang, C. W. Shu, and J. Ning, J. Comput. Phys., 231 (2012) 653.

\bibitem{Sun1995} J. Sun and J. Zhu, Theory of Detonation Physics, Beijing: National Defense Industry Press, 1995 (in Chinese).

\bibitem{Cochran1979} S. G. Cochran, J. Chan, UCID-18024, (1979).

\bibitem{LeeTarver1980} E. L. Lee and C. M. Tarver, Phys. Fluids, 23 (1980) 2362.

\bibitem{Cao1986} J. Cao, Explosion and Shock Waves, 6 (1986) 137 (in Chinese).

\bibitem{Zhao1989} F. Zhao, C. Sun, Y. Wei and J. Chi, Explosion and Shock Waves, 9 (1989) 338 (in Chinese).

\bibitem{Succi-Book} S. Succi, \textit{The Lattice Boltzmann Equation for Fluid Dynamics and Beyond}, Oxford University Press, New York, (2001).

\bibitem{Succi_RMP} S. Succi, I. V. Karlin, H. Chen, Rev. Mod. Phys. 74 (2002) 1203.

\bibitem{Succi_Sci} H. Chen, S. Kandasamy, S. Orszag, R. Shock, S. Succi, V. Yakhot, Science 301 (2003) 633.

\bibitem{Succi_PRL2005} A. Lamura, S. Succi, Phys. Rev. Lett. 95, 224502 (2005)

\bibitem{Succi_PRL2006A} M. Sbragaglia, R. Benzi, L. Biferale, S. Succi, and F. Toschi, Phys. Rev. Lett. 97 (2006) 204503.

\bibitem{Succi_PRL2006B} J. Horbach, S. Succi, Phys. Rev. Lett. 96 (2006) 224503.

\bibitem{Succi_PRL2009} R. Benzi, S. Chibbaro, S. Succi, Phys. Rev. Lett. 102 (2009) 026002.

\bibitem{Yeomans_PRL1997} G. Gonnella, E. Orlandini, and J. M. Yeomans, 78 (1997) 1695.

\bibitem{Yeomans_PRL2001} C. Denniston and J. M. Yeomans, Phys. Rev. Lett. 87 (2001) 275505.

\bibitem{Yeomans_PRL2002} G.Toth, C. Denniston, and J. M. Yeomans, Phys. Rev. Lett. 88 (2002) 105504.

\bibitem{Yeomans_PRL2004A} D. Marenduzzo, E. Orlandini, and J.M. Yeomans, 92 (2004) 188301.

\bibitem{Yeomans_PRL2004B} R. Verberg, C.M. Pooley, J.M. Yeomans, and A. C. Balazs, Phys. Rev. Lett. 93 (2004) 184501.

\bibitem{Yeomans_PRL2004C} C.M. Pooley and J.M. Yeomans, Phys. Rev. Lett. 93 (2004) 118001.

\bibitem{Yeomans_PRL2007} D. Marenduzzo, E. Orlandini, and J. M. Yeomans, Phys. Rev. Lett.  98 (2007) 118102.

\bibitem{Yeomans_PRL2013} A. Sengupta, U. Tkalec,  M. Ravnik, J. M. Yeomans, C. Bahr, and S. Herminghaus, Phys. Rev. Lett. 110 (2013) 048303.

\bibitem{ShanChen} X. W. Shan, H. D. Chen, Phys. Rev. E, 47 (1993) 1815; Phys. Rev. E, 49 (1994) 2941

\bibitem{SYChen} S. Chen, G. D. Doolen, Annu. Rev. Fluid Mech., 30 (1998) 329.

\bibitem{DXZhang} Q. Kang, D. Zhang, S. Chen, X. He, Phys. Rev. E, 65 (2002) 036318

\bibitem{HPFang} H. Fang, Z. Wang, Z. Lin, M. Liu, Phys. Rev. E, 65 (2002) 051925

\bibitem{Guozhaoli2013} Z. Guo, C. Shu, \textit{ Lattice Boltzmann Method and Its Applications in Engineering},  World Scientific Publishing Company, (2013).

\bibitem{ProgPhys2014} A. Xu, G. Zhang, Y. Li and H. Li, Prog. Phys. 34 (2014) 136 (in Chinese).

\bibitem{Dawson1993} S. Ponce Dawson, S. Chen, and G. D. Doolen. J. Chem. Phys., 98 (1993) 1514.

\bibitem{Weimar1996} J. R. Weimar, J. P. Boon. Physica A, 224 (1996) 207.

\bibitem{ZhangRenliang2014} R. Zhang, Y, Xu, B. Wen, et al, Scientific Reports, (2014) 4.

\bibitem{ChenShiyi1996} S. Chen, D. Martinez, R. Mei, Phys. Fluids, 8 (1996) 2527.

\bibitem{Succi1997} S. Succi, G. Bella, and F. Papetti, J. Sci. Comput., 12 (1997) 395

\bibitem{Filippova1998} O. Filippova, D H\"{a}nel, Int. J. Mod. Phys. C, 9 (1998) 1439.

\bibitem{Filippova2000JCP} O. Filippova, D H\"{a}nel, J. Comput. Phys, 158 (2000) 139.

\bibitem{Filippova2000CPC} O. Filippova, D H\"{a}nel, Comput. Phys. Commun., 129 (2000) 267

\bibitem{Yu2002} H. Yu, L. S. Luo, S. S. Girimaji, Int. J. Comput. Eng. Sci., 3 (2002) 73.

\bibitem{Yamamoto2002} K. Yamamoto, X. He, G. D. Doolen. J. Stat. Phys., 107 (2002) 367.

\bibitem{Yamamoto2003} K. Yamamoto, Int. J. Mod. Phys. B, 2003, 17(01n02): 197-200.

\bibitem{Yamamoto2005} K. Yamamoto, N. Takada, M. Misawa, P. Combust. Inst., 30 (2005) 1509.

\bibitem{Lee2006} T. Lee, C. Lin, and L. D. Chen, J. Comput. Phys., 215 (2006) 133.

\bibitem{Chiavazzo2010} E. Chiavazzo, I. V. Karlin, A. N. Gorban, et al., Combust. Flame, 157 (2010) 1833.

\bibitem{ChenSheng2007} S. Chen, Z. Liu, C. Zhang, et al., Appl. Math. Comput., 193 (2007) 266.

\bibitem{ChenSheng2008} S. Chen, Z. Liu, Z. Tian, et al, Comput. Math. Appl., 55 (2008) 1424.

\bibitem{ChenSheng2009} S. Chen, M. Krafczyk, Int. J. Therm. Sci., 48 (2009) 1978.

\bibitem{ChenSheng2010I} S. Chen, Int. J. Hydrogen Energ., 35 (2010) 1401.

\bibitem{ChenSheng2010II} S. Chen, J. Li, H. Han, et al., Int. J. Hydrogen Energ., 35 (2010) 3891.

\bibitem{ChenSheng2010III} S. Chen, H. Han, Z. Liu, et al., Int. J. Hydrogen Energ., 35 (2010) 4736.

\bibitem{ChenSheng2011} S. Chen, C. Zheng, Int. J. Hydrogen Energ., 36 (2011) 15403.

\bibitem{ChenSheng2012} S. Chen, J. Mi, H. Liu, et al., Int. J. of Hydrogen Energ., 37 (2012) 5234.

\bibitem{Alexander1992} F. J. Alexander, H. Chen, S. Chen, et al. Phys. Rev. A, 46 (1992) 1967.

\bibitem{Alexander1993} F. J. Alexander, S. Chen, J. D. Sterling. Phys. Rev. E, 47 (1993) R2249.

\bibitem{Chen1994} Y. Chen, H. Ohashi, M. Akiyama. Phys. Rev. E, 50 (1994) 2776.

\bibitem{McNamara1997} G. R. McNamara, A. L. Garcia, B. J. Alder. J. stat. phys., 87 (1997) 1111.

\bibitem{XuPan2007} X. Pan, A. Xu, G. Zhang, and S. Jiang, Int. J. Mod. Phys. C 18 (2007) 1747.

\bibitem{XuGan} Y. Gan, A. Xu, G. Zhang, and Y. Li, Physica A 387 (2008) 1721; Phys. Rev. E 83 (2011) 056704.

\bibitem{XuChen} F. Chen, A. Xu, G.Zhang, Y. Li, S. Succi, EuroPhys. Lett. 90 (2010) 54003.

\bibitem{Review2012} A. Xu, G. Zhang, Y. Gan, F. Chen, and X. Yu, Front. Phys. 7 (2012) 582.

\bibitem{XuYan2013} B. Yan, A. Xu, G. Zhang, Y. Ying, and H. Li, Frontiers of Physics, 8 (2013) 94.

\bibitem{Halliday2001} I. Halliday, L. A. Hammond, C. M. Care, et al. Phys. Rev. E, 64 (2001) 011208.

\bibitem{Niu2003} X. D. Niu, C. Shu, and Y. T. Chen, Int. J. Mod. Phys. C 14 (2003) 785.

\bibitem{Premnath2005} K. N. Premnath, J. Abraham. Phys. Rev. E, 71 (2005) 056706.

\bibitem{Reis2007} T. Reis and T. N. Phillips. Phys. Rev. E, 75 (2007) 056703.

\bibitem{Guo2009} Z. Guo, H. Han, B. Shi, et al. Phys. Rev. E, 79 (2009) 046708.

\bibitem{Watari2011} M. Watari, Commun. Comput. Phys. 9 (2011) 1293.

\bibitem{XuLin2014} C. Lin, A. Xu, G. Zhang, Y. Li, S. Succi, Phys. Rev. E, 89 (2014) 013307.

\bibitem{XuGan2013} Y. Gan, A. Xu, G. Zhang, Y. Yang, EPL, 103 (2013) 24003.

\bibitem{XuChen2013} F. Chen, A. Xu, G. Zhang, Y. Wang, Front. Phys. 9 (2014) 246.

\bibitem{Watari2003} M. Watari and M. Tsutahara, Phys. Rev. E 67 (2003) 036306.

\bibitem{Toro2009} E. F. Toro. Riemann solvers and numerical methods for
fluid dynamics: a practical introduction[M]. Springer, 2009.

\bibitem{WB1976} R. F. Warming, R. M. Beam. AIAA Journal,, 14 (1976) 1241.

\bibitem{XuLin_multi_distribution} C. Lin, A. Xu, G. Zhang, Y. Li, e-print arXiv: 1405.5500
\end{thebibliography}
\end{document}